%% file: main.tex
\documentclass{article}
\usepackage{spconfa4,amsmath,graphicx}
\usepackage{amsmath}
\usepackage{graphicx}
\usepackage{amssymb}
\usepackage{booktabs}
\usepackage{subcaption}
\usepackage{pifont}
\usepackage{url}
\usepackage{tikz}
\usepackage{pgfplots}

\newcommand{\cmark}{\ding{51}}
\newcommand{\xmark}{\ding{55}}

\title{Efficient Area-based and Speaker-Agnostic Source Separation}
\twoauthors
  {Martin Strauss\sthanks{Work done while the first author was doing an internship at Microsoft Applied Sciences Group.}}
	{International Audio Laboratories Erlangen\sthanks{A joint institution of the Friedrich-Alexander-Universität Erlangen-Nürnberg (FAU) and Fraunhofer IIS, Germany.}\\ Am Wolfsmantel 33, 91058 Erlangen, Germany\\martin.strauss@audiolabs-erlangen.de}
  {Okan Köpüklü}
	{Microsoft\\ Munich, Germany\\ okan.kopuklu@microsoft.com}
\begin{document}
\ninept
\maketitle
\begin{abstract}
        This paper introduces an area-based source separation method designed for virtual meeting scenarios. The aim is to preserve speech signals from an unspecified number of sources within a defined spatial area in front of a linear microphone array, while suppressing all other sounds. Therefore, we employ an efficient neural network architecture adapted for multi-channel input to encompass the predefined target area. To evaluate the approach, training data and specific test scenarios including multiple target and interfering speakers, as well as background noise are simulated. All models are rated according to DNSMOS and scale-invariant signal-to-distortion ratio. Our experiments show that the proposed method separates speech from multiple speakers within the target area well, besides being of very low complexity, intended for real-time processing. In addition, a power reduction heatmap is used to demonstrate the networks' ability to identify sources located within the target area. We put our approach in context with a well-established baseline for speaker-speaker separation and discuss its strengths and challenges.
\end{abstract}
\section{Introduction}
\label{sec:intro}
    In today's workplace, virtual meetings are an integral part even in an open office setting. These meetings may involve multiple individuals located in front of a single device like a laptop or a speakerphone. This setup faces a broad range of challenges with the number and location of participants in front of the device changing dynamically, but also disruptions from interfering speakers or background noise. Additionally, potential privacy concerns arise from capturing speech from individuals who are not actively participating in the meeting.

    A potential solution requires two main factors: (i) speech from all meeting participants need to be retained within a defined spatial area and (ii) the approach needs to be lightweight with respect to computational resources to avoid unwanted delay.

    Sound source separation techniques~\cite{speech_sep_book} are one possibility to separate the speech of meeting participants from all interfering sounds.  Many modern devices, such as headphones, smart-speakers or laptops, make use of more than one microphone to acquire the input audio mixture. This allows them to utilize spatial information, such as the location of individual sources, which is encoded into attenuation and time differences of arrival with respect to the microphone array. 
    
    Traditional spatial processing uses so-called beamformers, designed to enhance a signal arriving from a target direction while suppressing interfering sounds~\cite{souden2009optimal}. 
    Modern methods usually employ DNNs, either alongside traditional beamformers~\cite{Tesch2023, guided_se_2023} or as standalone non-linear spatial filtering approaches~\cite{tesch_icassp, tesch_2024}. 
    
    Lately, there has been increasing interest in approaches that are able to separate sources within specific locations, e.g., given the target source distance~\cite{yiwere2019, Taherian2022, patterson22_interspeech}, the direction of arrival \cite{Taherian2022, cos} or predefined regions~\cite{xu_2022_pmlr, wechsler2023}.
    For instance, \cite{cos} considers an unknown number of speakers that are localized and separated simultaneously within an angular region using a binary search algorithm. In~\cite{wechsler2023}, all sources are assumed to be located in predefined regions simulating a car-like scenario with fixed seat positions. In contrast,~\cite{patterson22_interspeech} performs source separation within a distance threshold of a single microphone, only relying on acoustic cues implicitly contained in the data. 
    
    Even though these methods provide good performance, their implementation either is of high computational complexity (e.g.~\cite{cos,xu_2022_pmlr,wechsler2023}) making them not suitable for real-time applications, or they assume a single active speaker (e.g.~\cite{guided_se_2023,tesch_2024}).
    
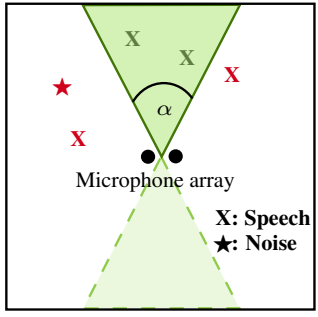
\begin{figure}[h]
        \centering
        \resizebox{0.25\textwidth}{!}{
            {\scalebox{1.05}{\input{setup.tikz}}}
    }
    \caption{Illustration of the investigated scenario. The speech sources inside the ROI with an angle of $\alpha=60^{\circ}$ are kept, while suppressing interfering speakers and noise. Speech sources and noise are denoted by \textbf{X} and $\bigstar$, respectively. The area and sources of interest are colorized in green. The dashed lines bound the mirrored area due to front-back ambiguity.} 
    \label{fig:setup}
    \end{figure}

    In this work, we propose an alternative approach. Our scenario assumes  one or multiple speakers attending a virtual meeting at the same time in front of a laptop with a two-microphone array. Additionally, interfering speech from other speakers present in the room and non-speech background noise are captured.
    Given this setup, the objective is to cover a pre-defined region-of-interest (ROI), which is independent from the specific location of individual speakers. This allows the DNN to preserve the speech of all individuals located within the ROI, even if multiple speakers are active at once. The ROI is defined by an angular span with the microphone array as origin (see Figure~\ref{fig:setup}). In addition, all remaining interfering sources are suppressed.
    This scenario is particularly challenging since the amount and location of speakers are unknown and not part of the training.
    
    We aim to solve this task in a data driven way, meaning that all the necessary information to cover the ROI should be entirely learned by implicit information contained in the data. Furthermore, to increase the applicability of our approach, we utilize an efficient DNN architecture capable of real-time processing. 

    The proposed scenario shares similarities to the ones in~\cite{rezero} and~\cite{yu2023deep}, but it includes more speakers and focuses on a two-microphone setup, as is common in conventional laptops.
    
    In the performance evaluation, we show that the proposed model is able to retain speech sources inside the ROI. At the same time it sufficiently suppresses interfering speakers and background noise. We also show that it outperforms a well-established speaker-speaker separation baseline in the most complex multi-target speaker scenario in terms of DNSMOS. Moreover, the proposed model offers lower complexity and faster processing speed. 

\section{Problem formulation}
\label{sec:problem}
    A uniform linear array (ULA) with $M \in \mathbb{N}^+$ microphones is placed at a random location in a room. The ROI is spanned by an angle $\alpha$ in front of the microphone array with its center as origin. 
    The setup is illustrated in Figure~\ref{fig:setup}. 
    For simplicity, the microphone array and all sources are assumed to be located at the same height. All reverberant time-domain speech sources, which are located inside the ROI and are captured by microphone $m$, are denoted as $\mathbf{t}_m\in \mathbb{R}^{N}$, i.e.,
	
	\begin{equation}
		\mathbf{t}_m = \sum_{i=1}^{I} \mathbf{t}_{m}^{i},
	\end{equation} 
	
\noindent where $i \in \left\{1,...,I\right\}$ is the speaker index and $N$ denotes the length of the signals in time samples. The microphone signals which are obtained from $j \in \left\{1,...,J\right\}$ interfering speakers outside the ROI are defined by $\textbf{k}_m \in \mathbb{R}^N$, i.e, 
	
	\begin{equation}
		\mathbf{k}_m = \sum_{j=1}^{J} \mathbf{k}_{m}^{j}.
	\end{equation}
	
	A single noise source $\textbf{n}_m \in \mathbb{R}^N$ can be located inside or outside the ROI. Consequently, the mixture $\textbf{y}_m \in \mathbb{R}^N$ which is captured by the $m^{th}$ microphone is the combination of all sources, i.e.,
	
	\begin{equation}
		\mathbf{y}_m = \mathbf{t}_{m} + \mathbf{k}_{m} + \mathbf{n}_m.
	\end{equation}
	
	The overall goal is to obtain a single-channel estimate $\hat{\mathbf{t}} \in \mathbb{R}^{N}$ of the summed speech sources located inside the ROI, while suppressing $\mathbf{k}$ and $\mathbf{n}$. 
 
\section{Proposed Methodology}
\label{sec:methods}	

    \subsection{Network architecture}
	We selected the CRUSE \cite{cruse} architecture due to its efficient design and real-time capability. It operates in the short-time Fourier transform (STFT) domain and applies a complex-valued single-channel mask $Q \in \mathbb{C}^{T\times F}$ to a complex-valued time-frequency (T-F) representation of the input signal, i.e.,
    
    \begin{equation}
        \hat{\mathbf{t}} = \text{iSTFT} \{ Q \odot \mathbf{Y}_{\phi}\},
    \end{equation}

    \noindent with $Y_{\phi} \in \mathbb{C}^{T\times F}$ being the STFT representation obtained by taking the average of all input channels of $\textbf{y}_m$. This choice was made in accordance to~\cite{Taherian2022_icassp}, in order to potentially extend this approach to other array geometries in the future. In the remainder of the paper, all variables with subscript $\phi$ were obtained the same way. We also ran experiments in using a single reference microphone to apply the mask and found only negligible performance differences.
    
    The original CRUSE architecture was developed for single-channel speech enhancement and trained on a complex compressed mean-squared error loss. To adapt the architecture for our task, we modified the initial convolution layer of the encoder to receive stereo audio as input. Therefore, the STFT representations of the left and right channel were concatenated. This way, the network should still be able to make use of the spatial information encoded in the multi-channel microphone data. 
    
    \begin{figure}[t]
        \centering
        \resizebox{0.48\textwidth}{!}{
            {\input{cruse.tikz}}
        }
    \caption{The network architecture. The input signal in STFT domain is concatenated along the microphone channel dimension and the output $Q$ is a single-channel complex-valued separation mask used to extract the target components.} 
    \label{fig:arch}
    \end{figure}
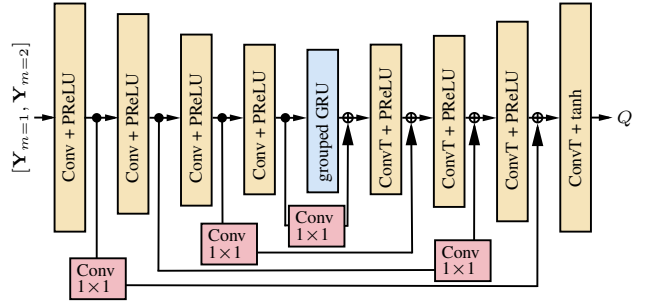

    The modified CRUSE architecture as illustrated in Figure~\ref{fig:arch} used 4 symmetric convolution and transposed convolution layers in the U-Net encoder and decoder blocks. A convolution kernel size of (2, 3) and stride of (1, 2) were used to down-sample and up-sample the frequency dimension at the convolution and transposed convolution layers, respectively. Skip connections were employed by using a trainable $1\times1$ convolution layer to add the corresponding encoder output to the decoder output, as was done in~\cite{cruse}. At the bottleneck layer, 4 parallel grouped Gated Recurrent Unit (GRU) layers~\cite{Ke2020} were used, which reduced the complexity and the number of parameters of the architecture. Each convolution and transposed convolution layer were followed by a PReLU activation function \cite{he2015delving}, except for the last transposed convolution layer, where a tanh activation function was used to produce the separation mask as output. 
    
    In order to investigate the effect of model complexity on the performance, we employed several variations of the model. $\textbf{CRUSE}_{l}$ and $\textbf{CRUSE}_{h}$ are models with light-weight (\textbf{l}) and heavy-weight (\textbf{h}) configurations, where the number of filters in the convolution layers were set to [32,\,64,\,64,\,64] and [32,\,64,\,128,\,256], respectively. 
	
\subsection{Training configuration}
\label{sec:conf}
	
    All the models were trained with an AdamW optimizer~\cite{loshchilov2018decoupled} (learning rate: $0.001$, weight decay: $2e-05$). For the STFT computation, a square-root Hann window of 20\,ms, hop size of 10\,ms and NFFT size of 320 points.
    As loss function, we employed the scale-invariant signal-to-distortion ratio (SI-SDR)~\cite{sisdr} between the target signal $\textbf{t}_{\phi}$ and the network output $\hat{\textbf{t}}$.

	
\section{Experimental Setup}
    \label{sec:exp}
    \subsection{Dataset}
    \label{sec:data}
    An offline synthetic dataset including a train, validation and test set was created for this study. The train and validation sets included speech and noise samples from the publicly available DNS-Challenge dataset \cite{dubey2023icassp}, ensuring not to mix up speakers for both sets. The \url{pyroomacoustics} \cite{pra} package was used to simulate virtual shoebox rooms of a randomly chosen size within [4.0\,m $\times$ 4.0\,m $\times$ 2.0\,m] to [8.0\,m $\times$ 8.0\,m $\times$ 4.0\,m]. T60 values typical for meeting rooms and offices \cite{speech_sep_book} were uniformly sampled at 0.25 -- 0.7\,s. The inter-microphone distance was 0.08\,m, with the array being placed randomly inside the room with at least 2\,m distance from each wall. The ROI was defined by an angle $\alpha=60^{\circ}$. 
    
    For this study, we created two versions of the dataset: A simple (\textbf{s}) version with 1 target and 1 interfering speaker per room, and a complex (\textbf{c}) version, with 1 to 4 uniformly sampled speakers placed inside and outside the ROI for each room. Consequently, one room contained at maximum 8 active speech sources. Subscripts of \textbf{s} and \textbf{c} denote the data setup used for training. To avoid problems with front-back ambiguity using ULAs, it was ensured to not place an interfering or target speech source in the mirrored target area along the microphone array. From each audio file, a 10\,s utterance was randomly extracted. In case the file was shorter than 10\,s, zero  padding to the desired length was employed. A sampling rate of $f_s=16$\,kHz was used for the data generation. The train and validation sets included approximately 55.6\,h and 22.2\,h of data, respectively.

    Target and interfering speakers were mixed according to a signal-to-interference ratio (SIR), which was uniformly sampled between 0 -- 10\,dB. Background noise was added to the speaker mix with a signal-to-noise ratio (SNR) values sampled from $\mathcal{N}(7,3)$, where $\mathcal{N}$ denotes a Gaussian distribution. All generated samples were level normalized with a value sampled at $\mathcal{N}(-28,10)$ dBFS.

    We evaluated performance using various test scenarios detailed in Table~\ref{tab:test_scens}. The speech files for the test data were taken from the 2020 Interspeech DNS-challenge~\cite{reddy20_interspeech} test set with noise from FSDnoisy18k~\cite{fsdnoisy}.  Scenarios 1 and 2 investigated the influence of different amounts of speakers with and without a noise. Additionally, different SIRs were investigated for a single target (scenario 3) and multi-target (scenario 4) setup. Each test scenario included 50 clips.

    \begin{table}[t]
    \centering
    \caption{Experimental setup of the various test scenarios with a mix of target \textbf{t} and interfering sources \textbf{k}. '\textit{random}' denotes random sampling concerning the number of speakers, source positions and SIRs.}
    \resizebox{0.42\textwidth}{!}{
    \begin{tabular}{@{}ccccc@{}}
    \toprule
    \textbf{Scen.} & \textbf{\#spk t} & \textbf{\#spk k} & \textbf{Noise} & \textbf{Setting} \\ \midrule
    \textbf{1} & 1 & 1 & \xmark,\cmark & random \\
    \textbf{2} & 2-4 & 1-4 & \xmark,\cmark & random \\ \midrule
    \textbf{3} & 1 & 1 &  \xmark & SIR: 0\,dB, 5\,dB, 10\,dB \\ 
    \textbf{4} &2-4 & 1-4 &  \xmark & SIR: 0\,dB, 5\,dB, 10\,dB \\ 
    \bottomrule
    \end{tabular}
    }
    \label{tab:test_scens}
    \end{table}

    \begin{table}[]
    \centering
    \caption{Comparison of employed architectures with respect to the number of parameters, computational complexity (GFLOPs) and real-time factor (RTF). The RTF numbers are the average processing time for 100 files of 10\,s length on a laptop with a 11th Gen. Intel(R) Core(TM) i7-1185G7 @ 3.00GHz.
    }
    \resizebox{0.39\textwidth}{!}{
    \begin{tabular}{@{}lccc@{}}
    \toprule
    \textbf{Model} & \multicolumn{1}{l}{\textbf{\# params {[}M{]}}} & \multicolumn{1}{l}{\textbf{\# GFLOPs}} & \textbf{RTF} \\ \midrule
    $\text{CRUSE}_{l}$ & 0.64 & \phantom{1 }9.18  & 0.04 \\
    $\text{CRUSE}_{h}$   & 8.58 & \phantom{ }38.20 & 0.07  \\
    Conv-TasNet & 5.08 & 112.34 & 0.24 \\ 
    \bottomrule
    \end{tabular}
    }
    \label{tab:comp}
    \end{table}

    \subsection{Performance evaluation}
    \noindent \textbf{Comparing method:} As direction of arrival estimation with multiple active speakers is challenging~\cite{8651493}, conventional beamformer baselines are inadequate for our needs.
    Therefore, as a well-established and available baseline, we trained a standard \textbf{Conv-TasNet}~\cite{convtasnet} on stereo input by increasing the input channel dimension in the encoder part of the network. The model applies a real-valued separation mask to a learned feature representation to estimate the combined utterances in the ROI. It was trained with the complex data setup described in Section~\ref{sec:data}.
    
    \noindent \textbf{Model complexity:} Table \ref{tab:comp} compares the employed architectures in terms of numbers of parameters, floating-point operation (FLOPs) and real-time factor (RTF). It can be seen that $\textbf{Conv-TasNet}$ is by far the most complex architecture compared to $\textbf{CRUSE}_{l}$ and $\textbf{CRUSE}_{h}$ in terms of GFLOPS and RTF. $\textbf{CRUSE}_{l}$ shows an approximately six times smaller amount of GFLOPS and two times lower processing time compared to the heavy-weight $\textbf{CRUSE}_{h}$ architecture. 
	
    \noindent \textbf{Computational metrics:} To evaluate the performance of the proposed approach, the SI-SDR and DNSMOS~\cite{DNSMOS} are used as performance metrics. DNSMOS is a non-intrusive quality metric used to estimate the outcome of a P.835 listening test~\cite{p835}. SI-SDR is a common metric to evaluate the signal quality, on which all models are also optimized via the loss function. We calculate the difference $\Delta$ of the computed metrics compared to the original input signal.
    
      \begin{figure}
        \centering
        \resizebox{0.39\textwidth}{!}{
        \includegraphics[width=\textwidth]{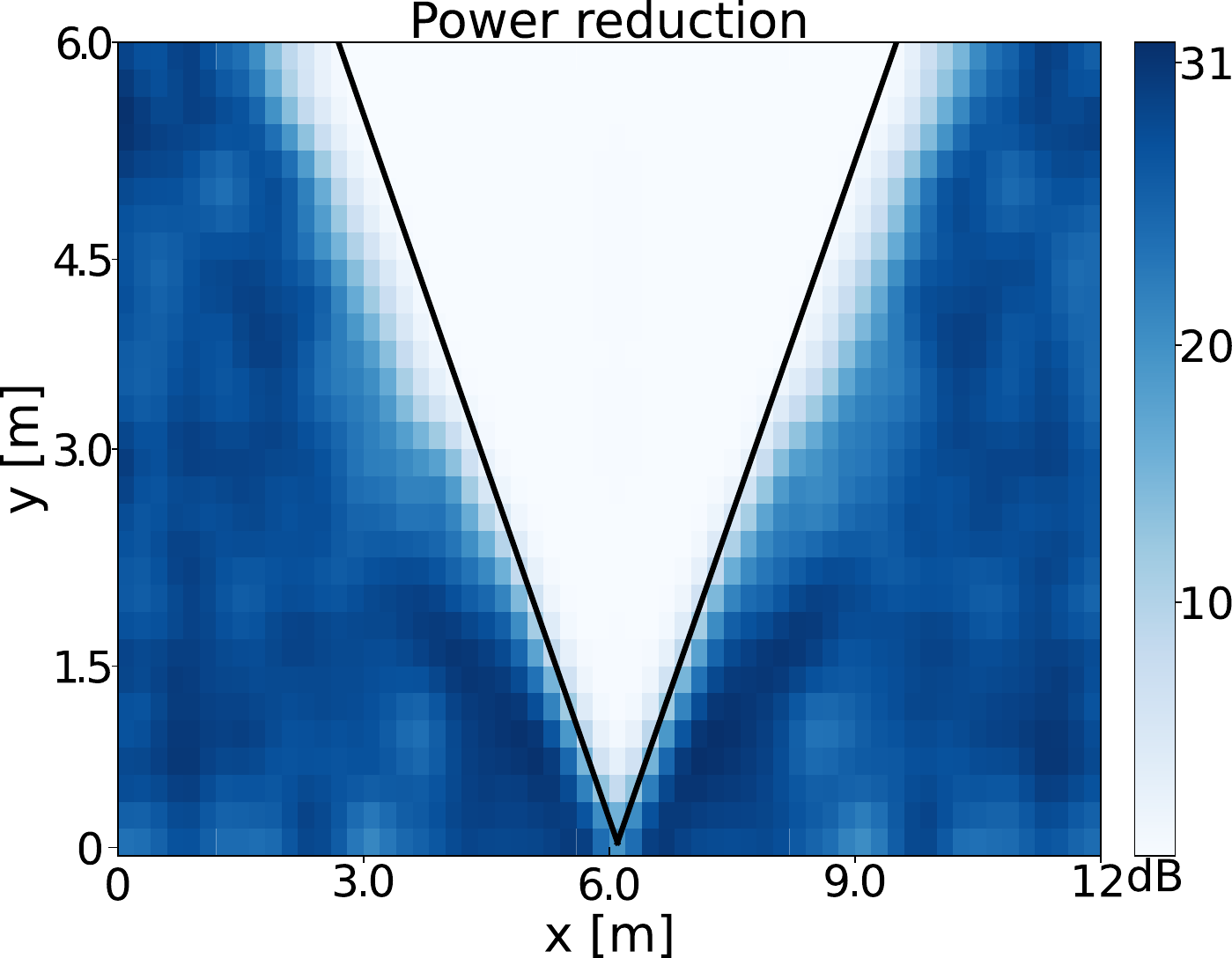}}
        \caption{PR heatmap of a ROI with $\alpha=60^{\circ}$ using the $\text{CRUSE}_{c,l}$ model. Due to front-back ambiguity for ULAs, only half the room is shown.}
        \label{fig:heatmap}
    \end{figure}

     \noindent \textbf{Power reduction (PR) heatmap:} To address the model's coverage of the entire ROI, we generated a so-called PR heatmap by placing a single speech source around a single room at an interval of 0.2\,m in x and y direction. Then, the PR metric from~\cite{patterson22_interspeech} was evaluated for each location, i.e.,
	
	\begin{equation}
		\text{PR}_{\text{dB}} := 10\log_{10} (||\textbf{y}_\phi||^2 / ||\hat{\textbf{t}}||^2).
	\end{equation}
 
    Ideally, if the source is located within the ROI, the model should output the unchanged signal with no PR. On the other hand, for each source located outside the ROI, we expect a strong PR  indicating a strong suppression of interfering sounds. The test room was set to a size of [12\,m $\times$ 12\,m $\times$ 2\,m], with a $T60=0.5$\,s and the microphone array was placed in the middle of the room.

    \begin{table*}[!htb]
     \centering
     \caption{DNSMOS and SI-SDR results (mean) for the test scenarios with a different number of speakers and with or without background noise. \textit{t} and \textit{k} represent target and interfering sources respectively. \textit{1} denotes that a single source was present, \textit{14} and \textit{24} represent randomly sampled 1 -- 4 and 2 -- 4 speakers, respectively.}
            \resizebox{\textwidth}{!}{
    \begin{tabular}{@{}l||cc|cc|cc|cc||cc|cc|cc|cc@{}}
    \toprule
     & \multicolumn{8}{c||}{\textbf{without noise}} & \multicolumn{8}{c}{\textbf{with noise}}   \\ \midrule
      & \multicolumn{2}{c|}{$\Delta$SIG}  & \multicolumn{2}{c|}{$\Delta$BAK} & \multicolumn{2}{c|}{$\Delta$OVRL}  & \multicolumn{2}{c||}{$\Delta$SI-SDR}  & \multicolumn{2}{c|}{$\Delta$SIG} & \multicolumn{2}{c|}{$\Delta$BAK}                & \multicolumn{2}{c|}{$\Delta$OVRL}   & \multicolumn{2}{c}{$\Delta$SI-SDR}  \\
     & \multicolumn{1}{c}{t1 k1} & \multicolumn{1}{c|}{t24 k14} & \multicolumn{1}{c}{t1 k1} & \multicolumn{1}{c|}{t24 k14} & \multicolumn{1}{c}{t1 k1} & \multicolumn{1}{c|}{t24 k14} & \multicolumn{1}{c}{t1 k1} & \multicolumn{1}{c||}{t24 k14} & \multicolumn{1}{c}{t1 k1} & \multicolumn{1}{c|}{t24 k14} & \multicolumn{1}{c}{t1 k1} & \multicolumn{1}{c|}{t24 k14} & \multicolumn{1}{c}{t1 k1} & \multicolumn{1}{c|}{t24 k14} & \multicolumn{1}{c}{t1 k1} & \multicolumn{1}{c}{t24 k14} \\ \midrule
    Conv-TasNet &  \textbf{0.03}  & 0.63   & 1.40 & 0.69 & \textbf{0.64} & 0.38 & \textbf{6.78}  & \textbf{3.16}  & \textbf{0.74} & 0.74 & 1.78  & 0.82 & \textbf{0.84} & 0.48 & \textbf{6.86} & \textbf{4.54} \\ \midrule
    $\text{CRUSE}_{s,l}$  & -0.05 & 0.61 &  \textbf{1.43} & 1.44  & 0.58 & 0.54 & 5.68 & 0.74 & 0.56 & 0.86 & 1.72 & 1.56 & 0.68 & 0.64 & 5.32 & 2.60 \\ 
    $\text{CRUSE}_{s,h}$  & 0.01 & \textbf{0.72} &  1.35 & \textbf{1.48}  & 0.59 & \textbf{0.57} & 6.24 & 1.11 & 0.68 & \textbf{0.97} & \textbf{1.80} & \textbf{1.66} & 0.77 & \textbf{0.69} & 5.83 & 3.03 \\ 
    $\text{CRUSE}_{c,l}$  & -0.10  & 0.61 & 1.21 & 0.66 & 0.45 & 0.36 & 5.31 & 2.69  & 0.62 & 0.66 & 1.39  & 0.70  & 0.61 & 0.41 & 5.10 & 3.37  \\
    $\text{CRUSE}_{c,h}$  &  -0.02 & 0.48 & 1.30 & 0.59 & 0.55 & 0.30  & 6.15  & 2.80  & 0.62 & 0.57 & 1.55 & 0.73 & 0.67 & 0.38  & 5.31 & 3.81  \\ \bottomrule
    \end{tabular}
    }
    \label{tab:spk_results}
    \end{table*}

    \begin{table}[!htb]
        \centering
        \caption{$\Delta$OVRL (mean $\pm$ std) of DNSMOS for different SIRs using 1 target and 1 interfering speaker without noise source.}
        \resizebox{0.4\textwidth}{!}{
        \begin{tabular}{@{}l|ccc@{}}
            \toprule
            & \multicolumn{3}{c}{SIR} \\  &0\,dB & 5\,dB & 10\,dB \\\midrule
            $\text{Conv-TasNet}$  & \textbf{0.43 $\pm$ 0.39} & \textbf{0.65 $\pm$ 0.44} & \textbf{0.68 $\pm$ 0.43} \\ \midrule
             $\text{CRUSE}_{s,l}$ & 0.36 $\pm$ 0.37 & 0.57 $\pm$ 0.44 & 0.65 $\pm$ 0.46\\
             $\text{CRUSE}_{s,h}$ & 0.40 $\pm$ 0.39 & 0.60 $\pm$ 0.41 & \textbf{0.68 $\pm$ 0.47}\\
            $\text{CRUSE}_{c,l}$ & 0.20 $\pm$ 0.34 & 0.46 $\pm$ 0.42 & 0.58 $\pm$ 0.45\\
        $\text{CRUSE}_{c,h}$ & 0.29 $\pm$ 0.34 & 0.52 $\pm$ 0.43 & 0.62 $\pm$ 0.45\\ 
            \bottomrule
        \end{tabular}
        }
        \label{tab:sir_res}
    \end{table}

    \begin{table}[!htb]
        \centering
	\caption{$\Delta$OVRL (mean $\pm$ std) of DNSMOS for different SIRs using multiple target and interfering speaker without noise source.}
 \resizebox{0.4\textwidth}{!}{
	\begin{tabular}{@{}l|ccc@{}}
			\toprule
			& \multicolumn{3}{c}{SIR} \\  &0\,dB & 5\,dB & 10\,dB \\\midrule
            $\text{Conv-TasNet}$  & 0.29 $\pm$ 0.39 & 0.42 $\pm$ 0.39 & 0.38 $\pm$ 0.36 \\ \midrule
		$\text{CRUSE}_{s,l}$ & 0.56 $\pm$ 0.25 & 0.61 $\pm$ 0.35 & 0.60 $\pm$ 0.30\\
  		$\text{CRUSE}_{s,h}$ & \textbf{0.62 $\pm$ 0.29} & \textbf{0.63 $\pm$ 0.34} & \textbf{0.66 $\pm$ 0.29}\\
            $\text{CRUSE}_{c,l}$ & 0.17 $\pm$ 0.32 & 0.37 $\pm$ 0.35 & 0.32 $\pm$ 0.32\\
		$\text{CRUSE}_{c,h}$ & 0.26 $\pm$ 0.33 & 0.34 $\pm$ 0.35 & 0.40 $\pm$ 0.35\\ 
            \bottomrule
    \end{tabular}
    }
    \label{tab:sir_res_multi_spk}
    \end{table}

\section{Results and Discussion}
\label{sec:res}

 \subsection{PR heatmap}
 Figure~\ref{fig:heatmap} displays the PR heatmap using the \textbf{$\text{CRUSE}_{c,l}$} model. The figure shows that the model is able to differentiate well between target area with almost no PR and the area outside the ROI. The effect is slightly reduced with further distance to the microphone array, where sources close to the edge of the ROI receive less suppression. It is worth mentioning that the heatmap visualization uses a larger room size that was never seen in training, suggesting that this method can generalize well to larger setups.
 
 \subsection{Varying amount of speakers}
 The results of test scenario 1 and 2 are displayed in Table~\ref{tab:spk_results}. Comparing the different CRUSE setups, it can be seen that \textbf{$\text{CRUSE}_{s,h}$} performs best in terms of DNSMOS 
 in most scenarios.
 This results suggest that it is sufficient to use a simple data setup to cover the target area for separation. The SI-SDR value of models with a simple data setup drops in the multi-target scenario without noise, showing a discrepancy between this metric and the DNSMOS values. Inspecting the corresponding test samples we found that those two models struggle to retain all target utterances for a small amount of test items where multiple target speakers are included, leading to a negative delta and a lower average value overall. \textbf{$\text{CRUSE}_{c,h}$} and \textbf{$\text{CRUSE}_{s,h}$} perform slightly better than their corresponding light-weight version in most settings. This indicates that a larger model can reach better values, at the cost of higher computational complexity.
 
 Compared to \textbf{Conv-TasNet}, it can be seen that  \textbf{$\text{CRUSE}_{s,l}$} and \textbf{$\text{CRUSE}_{s,h}$} perform better in a multi-target speaker setting in terms of the DNSMOS metric, while \textbf{Conv-TasNet} shows slightly better results for a single target speaker. However, it is worth emphasizing that \textbf{Conv-TasNet} is the most complex model, which operates on more than ten times the amount of computational operations. Although \textbf{Conv-TasNet} shows lower DNSMOS values in the multi-speaker setup, it performs best in terms of SI-SDR in all settings. Inspecting a few of the respective separated items reveal strong distortion-like artifacts in the generated samples for \textbf{Conv-TasNet}, which could explain the low DNSMOS values in part. Overall, considering perceptual quality and computational complexity the obtained results suggest \textbf{$\text{CRUSE}_{s,h}$} to be the best trade-off model.

\subsection{Varying SIRs}
 \noindent The results for different SIRs in test scenario 3 and 4 are displayed in Table \ref{tab:sir_res} and Table \ref{tab:sir_res_multi_spk}. Due to space constraints, only the $\Delta$OVRL of DNSMOS is reported here, however the overall trend remains the same for the other DNSMOS metrics. Both tables demonstrate that an increasing SIR leads to an increased performance, which is expected since the utterances placed inside the ROI become more dominant in the mixture. Looking at the multi-target scenario at 10\,dB no further gain is achieved compared to 5\,dB for \textbf{Conv-TasNet} and the light-weight CRUSE models. This suggests that in a complex scenario of several target and several interfering speakers, the effect of a higher SIR decreases.

 Overall, the results displayed in Table \ref{tab:sir_res} and Table \ref{tab:sir_res_multi_spk} largely confirm the results from Table \ref{tab:spk_results} with \textbf{Conv-TasNet} showing the strongest results with a single target speaker, but \textbf{$\text{CRUSE}_{s,h}$} being best in a multi-target speaker setup.
	
\section{Conclusion}
\label{sec:conclusion}
    This paper presented a real-time processing approach for area-based sound source separation in order to extract an unknown number of target speech utterances from an angular area in front of a linear microphone array. It was shown that the chosen approach is particularly strong in the setup where multiple sources are located inside the target area. In the future, we want to investigate the use of other microphone array setups and the possibility of an adaptable target area, depending on the application.

\bibliographystyle{IEEEbib}
\bibliography{refs}

\end{document}

%% file: setup.tikz
\tikzset{every picture/.style={line width=0.75pt}} 

\begin{tikzpicture}[x=0.75pt,y=0.75pt,yscale=-1,xscale=1]

\draw  [color={rgb, 255:red, 143; green, 203; blue, 75 }  ,draw opacity=1 ][fill={rgb, 255:red, 184; green, 233; blue, 134 }  ,fill opacity=0.28 ][dash pattern={on 4.5pt off 4.5pt}] (163.87,170.36) -- (196.92,234.89) -- (131,234.98) -- cycle ;
\draw  [fill={rgb, 255:red, 0; green, 0; blue, 0 }  ,fill opacity=1 ] (155.83,170.4) .. controls (155.83,169.02) and (156.95,167.9) .. (158.33,167.9) .. controls (159.71,167.9) and (160.83,169.02) .. (160.83,170.4) .. controls (160.83,171.78) and (159.71,172.9) .. (158.33,172.9) .. controls (156.95,172.9) and (155.83,171.78) .. (155.83,170.4) -- cycle ;
\draw  [fill={rgb, 255:red, 0; green, 0; blue, 0 }  ,fill opacity=1 ] (167.5,170.23) .. controls (167.5,168.85) and (168.62,167.73) .. (170,167.73) .. controls (171.38,167.73) and (172.5,168.85) .. (172.5,170.23) .. controls (172.5,171.61) and (171.38,172.73) .. (170,172.73) .. controls (168.62,172.73) and (167.5,171.61) .. (167.5,170.23) -- cycle ;
\draw  [color={rgb, 255:red, 65; green, 117; blue, 5 }  ,draw opacity=1 ][fill={rgb, 255:red, 184; green, 233; blue, 134 }  ,fill opacity=0.6 ] (164.09,170.37) -- (130.78,105.86) -- (197.15,105.74) -- cycle ;
\draw  [draw opacity=0] (151.88,147.09) .. controls (154.75,142.2) and (159.64,139.01) .. (165.15,139.07) .. controls (169.95,139.13) and (174.22,141.65) .. (177.06,145.58) -- (164.93,157.11) -- cycle ; \draw   (151.88,147.09) .. controls (154.75,142.2) and (159.64,139.01) .. (165.15,139.07) .. controls (169.95,139.13) and (174.22,141.65) .. (177.06,145.58) ;  
\draw   (98.48,105.29) -- (228.61,105.29) -- (228.61,235.42) -- (98.48,235.42) -- cycle ;

\draw (147.01,115.52) node [anchor=north west][inner sep=0.75pt]  [font=\scriptsize,color={rgb, 255:red, 95; green, 124; blue, 65 }  ,opacity=1 ] [align=left] {\textbf{X}};
\draw (189.01,130.86) node [anchor=north west][inner sep=0.75pt]  [font=\scriptsize,color={rgb, 255:red, 208; green, 2; blue, 27 }  ,opacity=1 ] [align=left] {\textbf{X}};
\draw (160.34,147.06) node [anchor=north west][inner sep=0.75pt]  [font=\scriptsize]  {$\alpha $};
\draw (126.34,175.37) node [anchor=north west][inner sep=0.75pt]  [font=\scriptsize] [align=left] {Microphone array};
\draw (185.34,191.52) node [anchor=north west][inner sep=0.75pt]  [font=\scriptsize] [align=left] {\textbf{X: Speech}};
\draw (170.01,123.52) node [anchor=north west][inner sep=0.75pt]  [font=\scriptsize,color={rgb, 255:red, 95; green, 124; blue, 65 }  ,opacity=1 ] [align=left] {\textbf{X}};
\draw (124.01,157.86) node [anchor=north west][inner sep=0.75pt]  [font=\scriptsize,color={rgb, 255:red, 208; green, 2; blue, 27 }  ,opacity=1 ] [align=left] {\textbf{X}};
\draw (116,135.4) node [anchor=north west][inner sep=0.75pt]  [font=\scriptsize,color={rgb, 255:red, 208; green, 2; blue, 27 }  ,opacity=1 ]  {$\bigstar $};
\draw (192.01,202.19) node [anchor=north west][inner sep=0.75pt]  [font=\scriptsize] [align=left] {\textbf{: Noise}};
\draw (184,202.07) node [anchor=north west][inner sep=0.75pt]  [font=\scriptsize]  {$\bigstar $};

\end{tikzpicture}

%% file: cruse.tikz
\tikzset{every picture/.style={line width=0.75pt}} 

\begin{tikzpicture}[x=0.75pt,y=0.75pt,yscale=-1,xscale=1]

\draw  [fill={rgb, 255:red, 246; green, 228; blue, 186 }  ,fill opacity=1 ] (100,40) -- (116,40) -- (116,160) -- (100,160) -- cycle ;
\draw  [fill={rgb, 255:red, 246; green, 228; blue, 186 }  ,fill opacity=1 ] (133,45.56) -- (149,45.56) -- (149,150.56) -- (133,150.56) -- cycle ;
\draw  [fill={rgb, 255:red, 246; green, 228; blue, 186 }  ,fill opacity=1 ] (166,56.33) -- (182,56.33) -- (182,146.33) -- (166,146.33) -- cycle ;
\draw  [fill={rgb, 255:red, 246; green, 228; blue, 186 }  ,fill opacity=1 ] (199,61.67) -- (215,61.67) -- (215,140.37) -- (199,140.37) -- cycle ;
\draw  [fill={rgb, 255:red, 211; green, 229; blue, 248 }  ,fill opacity=1 ] (232,64.33) -- (248,64.33) -- (248,139.33) -- (232,139.33) -- cycle ;
\draw  [fill={rgb, 255:red, 246; green, 228; blue, 186 }  ,fill opacity=1 ] (265,61.61) -- (281,61.61) -- (281,140.33) -- (265,140.33) -- cycle ;
\draw  [fill={rgb, 255:red, 246; green, 228; blue, 186 }  ,fill opacity=1 ] (298,57) -- (314,57) -- (314,147) -- (298,147) -- cycle ;
\draw  [fill={rgb, 255:red, 246; green, 228; blue, 186 }  ,fill opacity=1 ] (331,49.56) -- (347,49.56) -- (347,154.56) -- (331,154.56) -- cycle ;
\draw  [fill={rgb, 255:red, 246; green, 228; blue, 186 }  ,fill opacity=1 ] (364,40) -- (380,40) -- (380,160) -- (364,160) -- cycle ;
\draw    (89.38,100) -- (98.71,100) ;
\draw [shift={(100.71,100)}, rotate = 180] [fill={rgb, 255:red, 0; green, 0; blue, 0 }  ][line width=0.08]  [draw opacity=0] (7.2,-1.8) -- (0,0) -- (7.2,1.8) -- cycle    ;
\draw    (380,100) -- (389.33,100) ;
\draw [shift={(391.33,100)}, rotate = 180] [fill={rgb, 255:red, 0; green, 0; blue, 0 }  ][line width=0.08]  [draw opacity=0] (7.2,-1.8) -- (0,0) -- (7.2,1.8) -- cycle    ;
\draw   (250.67,100.06) .. controls (250.67,98.67) and (251.79,97.56) .. (253.17,97.56) .. controls (254.55,97.56) and (255.67,98.67) .. (255.67,100.06) .. controls (255.67,101.44) and (254.55,102.56) .. (253.17,102.56) .. controls (251.79,102.56) and (250.67,101.44) .. (250.67,100.06) -- cycle ; \draw   (250.67,100.06) -- (255.67,100.06) ; \draw   (253.17,97.56) -- (253.17,102.56) ;
\draw    (115.51,100.09) -- (131.56,100.09) ;
\draw [shift={(133.56,100.09)}, rotate = 180] [fill={rgb, 255:red, 0; green, 0; blue, 0 }  ][line width=0.08]  [draw opacity=0] (7.2,-1.8) -- (0,0) -- (7.2,1.8) -- cycle    ;
\draw  [fill={rgb, 255:red, 0; green, 0; blue, 0 }  ,fill opacity=1 ] (120,100.13) .. controls (120,99.02) and (120.9,98.13) .. (122,98.13) .. controls (123.1,98.13) and (124,99.02) .. (124,100.13) .. controls (124,101.23) and (123.1,102.13) .. (122,102.13) .. controls (120.9,102.13) and (120,101.23) .. (120,100.13) -- cycle ;
\draw    (149.51,100.09) -- (163.71,100.18) ;
\draw [shift={(165.71,100.19)}, rotate = 180.35] [fill={rgb, 255:red, 0; green, 0; blue, 0 }  ][line width=0.08]  [draw opacity=0] (7.2,-1.8) -- (0,0) -- (7.2,1.8) -- cycle    ;
\draw    (182.51,100.09) -- (196.71,100.18) ;
\draw [shift={(198.71,100.19)}, rotate = 180.35] [fill={rgb, 255:red, 0; green, 0; blue, 0 }  ][line width=0.08]  [draw opacity=0] (7.2,-1.8) -- (0,0) -- (7.2,1.8) -- cycle    ;
\draw  [fill={rgb, 255:red, 0; green, 0; blue, 0 }  ,fill opacity=1 ] (152.48,100.13) .. controls (152.48,99.02) and (153.37,98.13) .. (154.48,98.13) .. controls (155.58,98.13) and (156.48,99.02) .. (156.48,100.13) .. controls (156.48,101.23) and (155.58,102.13) .. (154.48,102.13) .. controls (153.37,102.13) and (152.48,101.23) .. (152.48,100.13) -- cycle ;
\draw  [fill={rgb, 255:red, 0; green, 0; blue, 0 }  ,fill opacity=1 ] (185.48,100.13) .. controls (185.48,99.02) and (186.37,98.13) .. (187.48,98.13) .. controls (188.58,98.13) and (189.48,99.02) .. (189.48,100.13) .. controls (189.48,101.23) and (188.58,102.13) .. (187.48,102.13) .. controls (186.37,102.13) and (185.48,101.23) .. (185.48,100.13) -- cycle ;
\draw    (215.51,100.09) -- (229.71,100.18) ;
\draw [shift={(231.71,100.19)}, rotate = 180.35] [fill={rgb, 255:red, 0; green, 0; blue, 0 }  ][line width=0.08]  [draw opacity=0] (7.2,-1.8) -- (0,0) -- (7.2,1.8) -- cycle    ;
\draw  [fill={rgb, 255:red, 0; green, 0; blue, 0 }  ,fill opacity=1 ] (218.48,100.13) .. controls (218.48,99.02) and (219.37,98.13) .. (220.48,98.13) .. controls (221.58,98.13) and (222.48,99.02) .. (222.48,100.13) .. controls (222.48,101.23) and (221.58,102.13) .. (220.48,102.13) .. controls (219.37,102.13) and (218.48,101.23) .. (218.48,100.13) -- cycle ;
\draw    (248.51,100.09) -- (262.71,100.18) ;
\draw [shift={(264.71,100.19)}, rotate = 180.35] [fill={rgb, 255:red, 0; green, 0; blue, 0 }  ][line width=0.08]  [draw opacity=0] (7.2,-1.8) -- (0,0) -- (7.2,1.8) -- cycle    ;
\draw   (283.67,100.06) .. controls (283.67,98.67) and (284.79,97.56) .. (286.17,97.56) .. controls (287.55,97.56) and (288.67,98.67) .. (288.67,100.06) .. controls (288.67,101.44) and (287.55,102.56) .. (286.17,102.56) .. controls (284.79,102.56) and (283.67,101.44) .. (283.67,100.06) -- cycle ; \draw   (283.67,100.06) -- (288.67,100.06) ; \draw   (286.17,97.56) -- (286.17,102.56) ;
\draw    (281.51,100.09) -- (295.71,100.18) ;
\draw [shift={(297.71,100.19)}, rotate = 180.35] [fill={rgb, 255:red, 0; green, 0; blue, 0 }  ][line width=0.08]  [draw opacity=0] (7.2,-1.8) -- (0,0) -- (7.2,1.8) -- cycle    ;
\draw   (316.67,100.06) .. controls (316.67,98.67) and (317.79,97.56) .. (319.17,97.56) .. controls (320.55,97.56) and (321.67,98.67) .. (321.67,100.06) .. controls (321.67,101.44) and (320.55,102.56) .. (319.17,102.56) .. controls (317.79,102.56) and (316.67,101.44) .. (316.67,100.06) -- cycle ; \draw   (316.67,100.06) -- (321.67,100.06) ; \draw   (319.17,97.56) -- (319.17,102.56) ;
\draw    (314.51,100.09) -- (328.71,100.18) ;
\draw [shift={(330.71,100.19)}, rotate = 180.35] [fill={rgb, 255:red, 0; green, 0; blue, 0 }  ][line width=0.08]  [draw opacity=0] (7.2,-1.8) -- (0,0) -- (7.2,1.8) -- cycle    ;
\draw   (349.67,100.06) .. controls (349.67,98.67) and (350.79,97.56) .. (352.17,97.56) .. controls (353.55,97.56) and (354.67,98.67) .. (354.67,100.06) .. controls (354.67,101.44) and (353.55,102.56) .. (352.17,102.56) .. controls (350.79,102.56) and (349.67,101.44) .. (349.67,100.06) -- cycle ; \draw   (349.67,100.06) -- (354.67,100.06) ; \draw   (352.17,97.56) -- (352.17,102.56) ;
\draw    (347.51,100.09) -- (361.71,100.18) ;
\draw [shift={(363.71,100.19)}, rotate = 180.35] [fill={rgb, 255:red, 0; green, 0; blue, 0 }  ][line width=0.08]  [draw opacity=0] (7.2,-1.8) -- (0,0) -- (7.2,1.8) -- cycle    ;
\draw    (122,100) -- (122,174.09) ;
\draw    (220.48,102.13) -- (220.43,157.18) ;
\draw    (253.16,104.56) -- (253.13,111.12) -- (252.93,156.68) ;
\draw [shift={(253.17,102.56)}, rotate = 90.25] [fill={rgb, 255:red, 0; green, 0; blue, 0 }  ][line width=0.08]  [draw opacity=0] (12,-3) -- (0,0) -- (12,3) -- cycle    ;
\draw    (187.48,102.13) -- (187.67,156.12) ;
\draw    (154.48,102.13) -- (154.21,180.59) ;
\draw    (286.18,104.56) -- (286.67,171.46) ;
\draw [shift={(286.17,102.56)}, rotate = 89.58] [fill={rgb, 255:red, 0; green, 0; blue, 0 }  ][line width=0.08]  [draw opacity=0] (12,-3) -- (0,0) -- (12,3) -- cycle    ;
\draw    (319.16,104.56) -- (319.01,164.99) ;
\draw [shift={(319.17,102.56)}, rotate = 90.14] [fill={rgb, 255:red, 0; green, 0; blue, 0 }  ][line width=0.08]  [draw opacity=0] (12,-3) -- (0,0) -- (12,3) -- cycle    ;
\draw    (352.16,105) -- (352.03,189.13) ;
\draw [shift={(352.17,103)}, rotate = 90.09] [fill={rgb, 255:red, 0; green, 0; blue, 0 }  ][line width=0.08]  [draw opacity=0] (12,-3) -- (0,0) -- (12,3) -- cycle    ;
\draw  [fill={rgb, 255:red, 242; green, 189; blue, 195 }  ,fill opacity=1 ] (176.67,155.78) -- (205.67,155.78) -- (205.67,177.28) -- (176.67,177.28) -- cycle ;

\draw  [fill={rgb, 255:red, 242; green, 189; blue, 195 }  ,fill opacity=1 ] (222.33,146.37) -- (251.33,146.37) -- (251.33,167.86) -- (222.33,167.86) -- cycle ;

\draw  [fill={rgb, 255:red, 242; green, 189; blue, 195 }  ,fill opacity=1 ] (108.17,174.03) -- (137.17,174.03) -- (137.17,195.53) -- (108.17,195.53) -- cycle ;

\draw  [fill={rgb, 255:red, 242; green, 189; blue, 195 }  ,fill opacity=1 ] (298.63,164.53) -- (327.63,164.53) -- (327.63,186.03) -- (298.63,186.03) -- cycle ;

\draw    (220.05,156.7) -- (222.41,156.7) ;
\draw    (251.05,156.97) -- (253.41,156.97) ;
\draw    (206,171) -- (286.67,171) ;
\draw    (153.7,180.59) -- (298.53,180.18) ;
\draw    (136.93,188.18) -- (352.03,188.68) ;

\draw (102.51,135.17) node [anchor=north west][inner sep=0.75pt]  [font=\scriptsize,rotate=-269.99] [align=left] {Conv + PReLU};
\draw (135.84,135.83) node [anchor=north west][inner sep=0.75pt]  [font=\scriptsize,rotate=-269.99] [align=left] {Conv + PReLU};
\draw (169,135.83) node [anchor=north west][inner sep=0.75pt]  [font=\scriptsize,rotate=-269.99] [align=left] {Conv + PReLU};
\draw (202,134.5) node [anchor=north west][inner sep=0.75pt]  [font=\scriptsize,rotate=-269.99] [align=left] {Conv + PReLU};
\draw (76.4,131.09) node [anchor=north west][inner sep=0.75pt]  [font=\scriptsize,rotate=-270]  {$[\mathbf{Y}_{m=1} ,\mathbf{Y}_{m=2}]$};
\draw (235,135.33) node [anchor=north west][inner sep=0.75pt]  [font=\scriptsize,rotate=-269.99] [align=left] {grouped GRU};
\draw (268,138.33) node [anchor=north west][inner sep=0.75pt]  [font=\scriptsize,rotate=-269.99] [align=left] {ConvT + PReLU};
\draw (301,139) node [anchor=north west][inner sep=0.75pt]  [font=\scriptsize,rotate=-269.99] [align=left] {ConvT + PReLU};
\draw (334,139.83) node [anchor=north west][inner sep=0.75pt]  [font=\scriptsize,rotate=-269.99] [align=left] {ConvT + PReLU};
\draw (367,138.17) node [anchor=north west][inner sep=0.75pt]  [font=\scriptsize,rotate=-269.99] [align=left] {ConvT + tanh};
\draw (392,95.16) node [anchor=north west][inner sep=0.75pt]  [font=\scriptsize]  {$Q$};
\draw (178.69,165.8) node [anchor=north west][inner sep=0.75pt]  [font=\scriptsize] [align=left] {1$\displaystyle \times $1};
\draw (178.67,157.16) node [anchor=north west][inner sep=0.75pt]  [font=\scriptsize] [align=left] {Conv\\};
\draw (224.33,147.74) node [anchor=north west][inner sep=0.75pt]  [font=\scriptsize] [align=left] {Conv\\};
\draw (224.35,156.38) node [anchor=north west][inner sep=0.75pt]  [font=\scriptsize] [align=left] {1$\displaystyle \times $1};
\draw (110.17,175.41) node [anchor=north west][inner sep=0.75pt]  [font=\scriptsize] [align=left] {Conv\\};
\draw (110.19,184.05) node [anchor=north west][inner sep=0.75pt]  [font=\scriptsize] [align=left] {1$\displaystyle \times $1};
\draw (300.63,165.91) node [anchor=north west][inner sep=0.75pt]  [font=\scriptsize] [align=left] {Conv\\};
\draw (300.65,174.55) node [anchor=north west][inner sep=0.75pt]  [font=\scriptsize] [align=left] {1$\displaystyle \times $1};

\end{tikzpicture}